\documentstyle[preprint,eqsecnum,aps,prb]{revtex}

\begin{document}
\title{Relations between structural distortions and transport properties in Nd$%
_{0.5}$Ca$_{0.5}$MnO$_{3}$ strained thin films.}
\author{E.\ Rauwel Buzin, W. Prellier\thanks{%
prellier@ismra.fr}, B.\ Mercey, Ch. Simon and B.\ Raveau}
\address{Laboratoire CRISMAT, CNRS\ UMR 6508, 6 Bd du\\
Mar\'{e}chal Juin, 14050 Caen Cedex, FRANCE.}
\date{\today}
\maketitle

\begin{abstract}
Strained thin films of charge/orbital ordered (CO/OO) $%
Nd_{0.5}Ca_{0.5}MnO_{3}$ (NCMO) with various thickness have grown on
(100)-SrTiO$_{3}$ and (100)-LaAlO$_{3}$ substrates, by using the Pulsed
Laser Deposition (PLD) technique. The thickness of the films influences
drastically the transport properties. As the thickness decreases, the CO
transition increases while at the same time the insulator-to-metal
transition temperature decreases under application of a 7T magnetic field.
Clear relationships between the structural distortions and the transport
properties are established. They are explained on the basis of the
elongation and the compression of the Mn-O-Mn and Mn-O bond angles and
distances of the {\it Pnma} structure, which modify the bandwidth and the
Jahn-Teller distortion in these materials.
\end{abstract}

\pacs{}

\newpage

\section{Introduction}

Perovskite type manganites such as R$_{1-x}$A$_x$MnO$_3$ (R=rare earth ion
and A=alkaline earth ion) have been extensively investigated due their
properties of colossal magnetoresistance (CMR) \cite{1}. The CMR effects
originates from a competition between a ferromagnetic metallic (FMM) state
and an antiferromagnetism insulating (AFMI) state. The appearance of the FMM
state is explained by the double exchange (DE) mechanism \cite{2}, whereas
the AFMI state originates from the Jahn-Teller (JT) distortions \cite
{Millis,3,4}, leading in the case of small A-site cations to the
charge/orbital ordered (CO/OO) phenomenon \cite{5}. The efforts made for
understanding the magnetotransport properties of manganite thin films \cite
{1,6,7} have shown that the physical properties are often different from the
bulk materials.\ The main reason deals with the strains included the
substrate-induced strains which modify the lattice parameters of the film
and change the properties as shown recently for $Pr_{0.5}Ca_{0.5}MnO_3$ \cite
{8,9,10}, $La_{1-x}Ba_xMnO_3$ \cite{11} and $Nd_{0.5}Ca_{0.5}MnO_3$ \cite{12}%
.

In manganite thin films, the strains are susceptible to modify the Mn-O-Mn
angles and Mn-O distances. Consequently, this affects the bandwidth and also
the JT distortions, so that DE and CO/OO\ phenomena may be drastically
influenced. Based on these considerations, the strain effects should
decrease as the thickness increases and it should be possible to establish
relationships between the structural distortions and the transport
properties, by studying the \ properties of the films with various
thicknesses.

Starting from $Nd_{0.5}Ca_{0.5}MnO_3$ films where a spectacular CMR\ effect
was previously evidenced \cite{12}, compared to the bulk \cite{13}, we have
carried out a systematic study of films with different thickness grown in
situ using the Pulsed Laser Deposition technique (PLD) on both (100)-SrTiO$%
_3 $ and (100)-LaAlO$_3$ substrates. In this paper, a clear correlation
between the structural distortions and the transport properties
(insulator-to-metal $T_{IM}$ transition under a magnetic field, CO/OO $T_{CO}
$ transition). We suggested that for small films thickness the elongation of
the Mn-O bonds prevails, reinforcing the JT\ distortion and consequently the
CO/OO state, whereas for large thickness the flattering of the Mn-O-Mn bond
angles prevails, increasing the bandwidth favoring the insulator-to-metal
transition to the detriment of the CO/OO state.

\section{Experimental}

Films with different thicknesses were deposited on single crystal substrates
of (100)-SrTiO$_3$ (STO) and (100)-LaAlO$_3$ (LAO) from a dense targets of Nd%
$_{0.5}$Ca$_{0.5}$MnO$_3$ using the PLD\ technique. Detailed of the process
can be found elsewhere \cite{12}. The structural study was done by X-ray
diffraction (XRD) using a Seifert XRD 3000P for the $\Theta $-2$\Theta $
scans and a Phillips MRD X'pert for the in-plane measurements (Cu K$\alpha $%
, $\lambda $=1.5406\AA ). The in-plane lattice parameters were obtained from
the (103)$_C$ reflection (where C refers to the ideal cubic perovskite cell, 
$a_C$=3.9\AA ). An electron microscope JEOL 2010 equipped with an Energy
Dispersive Spectroscopy (EDS) analysis, was used for the electron
diffraction (ED) study. The EDS has shown that the cationic composition of
the film is homogeneous and remains very close to the nominal one ''Nd$%
_{0.5\pm 0.02}$Ca$_{0.5\pm 0.02}$MnO$_x$'' within experimental error.

The resistivity ($\rho $) of the films was measured with a PPMS Quantum
Design as a function of the magnetic field ($H$) and the temperature ($T$)
using four-probe contacts. The thickness ($t$) of the films is measured
using a Dektak$^{3}$ST surface profiler.

Besides the classical growth parameters such as the deposition temperature,
the oxygen pressure, we found that the energy of the laser was also
important in order to obtain a single phase. The optimum energy value is $%
200mJ$ which corresponds to $2J/cm^2$ on the target. This is evidenced on
Fig.1 for a 2000\AA\ thick film of NCMO on STO. A low ($132mJ$) or a high ($%
240mJ$) laser energy leads to a poor crystallization of the film with
appearance of a secondary phase.

\section{Results}

\subsection{Structural properties}

The structure of bulk NCMO is orthorhombic ($Pnma$) with $a=5.4037$\AA , $%
b=7.5949$\AA\ and $c=5.3814$\AA\ \cite{14}. Some area of the film have been
investigated by ED and we found that the films are single phase,
[010]-oriented, i.e. with the [010]-axis perpendicular to the STO substrate
plane, and [101]-oriented, i.e. with the [101]-axis perpendicular to the LAO
substrate plane in the space group $Pnma$. This orientation is not
surprising and resulting from the lattice mismatch between the film and the
substrate as previously reported for Pr$_{0.5}$Ca$_{0.5}$MnO$_{3}$ thin
films grown on (100)-SrTiO$_{3}$ and (001)-LaAlO$_{3}$ substrates \cite{8,9}%
. Details of the transmission electron microscopy study are currently
undertaken and will be published elsewhere.

Fig.2a shows the evolution of the lattice parameters of the NCMO film grown
on STO versus thickness. One observed that the in-plane lattice parameter $%
d_{101}$ decreases as the thickness of the film increases while at the same
time, the out-of-plane lattice parameter $d_{020}$ increases. The lattice
parameters of films on LAO were also determined (Fig.2b) but, due to the
twins of this substrates, we were not able to obtain the in-plane lattice
parameters. The out-of-plane lattice parameters are almost constant on LAO
when the thickness is changing, suggesting that the film is quasi-relaxed
even for small thickness.

\subsection{Transport properties}

Figure 3a shows the resistivity dependence of the temperature for NCMO films
on STO. Without a magnetic field, all films present a semiconducting
behavior with an anomaly around 275K, corresponding to the $T_{CO}$. A
similar effect is observed on the bulk, but at a lower temperature close to
250K \cite{15}. In fact, this anomaly in the $\rho (T)$ curves is more clear
(inset of Fig. 3a), when the resistance is plotted in a logarithmic scale
versus the inverse of the temperature. When applying a magnetic field of 7T,
while the thinner film ($t=670$\AA ) remains insulating in the whole
temperature range of 4-300K, the thicker films ($t>1340$\AA ) show an
insulator-to-metal transition ($T_{IM}$) at 100K and 130K for the 1340\AA\
and 2000\AA\ films respectively. Such an effect is typical of the
metastability of the CO/OO state where a magnetic field (abbreviated $H_{C}$%
) can induce a metallic behavior at low temperature. However, the magnetic
field required is much higher in the bulk compound (close to 20T for the
same temperature) \cite{13}. In fact, there is a dependence of the critical
field with the temperature (Figure 3b). The resistivity shows an important
decrease on a logarithmic scale at a critical field ($H_{C}$) indicating the
field-induced melting of the CO/OO state. This field-induced
insulator-to-metal transition, which accompanies the collapsing of the CO/OO
state takes place below $T_{CO}$. This decrease can be viewed as a CMR of
about four orders of magnitude at 75K. A clear hysteresis (between the lower
and the upper critical fields) is seen at these temperatures as previously
reported on several CO/OO compounds \cite{5,16,17} but this hysteretic
region is more pronounced when the temperature is decreasing (see $\rho (H)$
at 25K in Fig.3b). In addition, the temperature dependence of the large
hysteresis region is a feature of a first order transition and has been
extensively studied for the composition Nd$_{0.5}$Sr$_{0.5}$MnO$_{3}$ in
Ref. \cite{16}. Note also the reentrant nature of the CO at 25K due to the
fact that a 7T magnetic field is not enough to completely melt this state.

On LAO substrates, the $\rho (T)$ curves are practically similar (Fig.4a) to
those on STO except that in this case, even the thinnest film (670\AA )
displays an insulator-to-metal transition when applying a magnetic field of
7T. Moreover, the $T_{IM}$ is higher than in the case of STO\ which is
consistent with the quasi-relaxed NCMO film found on LAO (for a 1340\AA\
film, $T_{IM}<100K$ on STO\ and $T_{IM}=110K$ on LAO). Large hystereses
indicating the first-order phase transition are also observed (Fig.4b) and,
in the same way, there is an increase of the $T_{CO}$ when the thickness of
the film decreases (inset of Fig. 4a).

\subsection{Discussion}

This study shows that the thickness of NCMO films deposited on STO\
influences strongly the transport properties of the material. For the
largest thickness ($t=2000$\AA ), an insulator-to-metal transition is
induced under 7T\ \ in contrast to the bulk \cite{13}, although the cell
parameters reach values close to those of the bulk ceramics \cite{14}.\
Moreover, the $T_{IM}$ under 7T magnetic field decreased as the thickness
decreases whereas at the same time, the $T_{CO}$ increases so that for a
small thickness ($t=650$\AA ) the film remains insulator.

Though it was suggested by several authors in bulk samples of the same
composition that electronic phase separation can be at the origin of the
drop in resistivity \cite{14,14a}, no evidence of such phase separation was
found in the thin films.\ We have indeed observed a continuous variation of
the lattice parameters with the thickness of the films.

Thus, in order to explain this evolution, we have take into consideration
the following features concerning the strains induced by the STO\ substrates

(i) The tensile stain along the [101] direction (in the plane of the
substrate) can increases both the Mn-O-Mn angles and Mn-O distances in the
equatorial plane (see Fig.5a). The increase of the Mn-O$_{EQ}$-Mn \cite{18}
bond angle increases the bandwith \cite{Rao}, the angles tending toward $180{%
{}^\circ}$, so that DE is involved. Consequently this structural effect
should destabilize the CO/OO\ state and favor the $T_{IM}$.\ The increase of
the Mn-O bond length induces the antagonist effect, favoring the distortion
of the MnO$_{6}$ octahedra, i.e. the Jahn-Teller distortion, so that the
CO/OO\ state should be stabilized.

(ii) The compression of the structure along the [010] direction (along the
out-of-plane direction on STO), which appears simultaneously, results in a
decrease of the Mn-O$_{AX}$-Mn \cite{18} angle along this direction.
Consequently, this decreases the bandwidth and stabilizes the CO/OO\ state.

(iii) In the $Pnma$ structure (such as NCMO), there are eight Mn-O$_{EQ}$-Mn
bonds against four Mn-O$_{AX}$-Mn bons.\ Thus, the bandwidth along the Mn-O$%
_{EQ}$ \ is consistently larger than the bandwidth along the Mn-O$_{AX}$
direction which indiquates that the DE along the [101] direction prevails
over the [010] direction \cite{Rao}.

(iv) Viewing these remarks, it is reasonable to assume that the
substrate-induced strains modify in a first step the tilting of the
octahedra, i.e. the Mn-O-Mn angles and only in a second step the Mn-O bond
lengths in the plane of the substrate.

For the larger films thickness ($t>2000$\AA ), the cell parameters are close
to those of the bulk leading consequently to a tilting of the octahedra
rather than a modification of the bond lengths. In other words, the
variation of the Mn-O-Mn angles prevails over the variation of the Mn-O
distances. Thus, as $d_{101}$ increases the Mn-O$_{EQ}$-Mn angles increases
with respect to the bulk. In contrast, as $d_{010}$ decreases the Mn-O$_{AX}$%
-Mn angle decreases, resulting in an increase of the bandwidth along the
[101] direction and a decrease of the bandwidth along the [010] direction.\
But as pointed out above, the first effect prevails over the second one. As
a consequence, the DE favored with respect to the bulk, so that the CO/OO\
state is less stable and an $T_{IM}$ is obtained under a 7T magnetic field
contrary to the bulk.

As the thickness of the film decreases from 2000\AA , $d_{101}$ increases
significantly so the Mn-O$_{EQ}$-Mn bond angle reached 180${{}^\circ}$
rapidly and the Mn-O$_{EQ}$ bond length increases. The increase of the Mn-O$%
_{EQ}$ distance induces a JT\ distortion of the octahedra, stabilizing the
CO/OO\ state at the detriment of the $T_{IM}$ transition. This explains why
the $T_{CO}$ transition increases as the thickness of the film decreases and
the $T_{IM}$ transition under 7T decreases correlatively. Finally, for the
smaller film thickness ($t=650$\AA ), the $d_{101}$ is very close to the
STO\ parameter, and the bond angle Mn-O$_{EQ}$ effects prevails over the Mn-O%
$_{EQ}$-Mn effect, so that the CO/OO state is the most stable. Note that the
simultaneous decrease of the $d_{010}$, i.e. decrease of Mn-O$_{AX}$-Mn also
favors the stabilization of the CO/OO\ state.

In the case of NCMO films grown on LAO, the substrate-induced strains lead
in fact to rather similar effects.\ Indeed, one observes a compression along
the [010]-axis in the plane of the substrate. As shown in Fig.5b, this
induces a decreases of the Mn-O$_{AX}$-Mn angles and of the Mn-O$_{AX}$ bond
lengths similar to what occurs for STO; an increase of the Mn-O$_{EQ}$-Mn
out-of-plane bond angle is induced, but in this case the increase of the Mn-O%
$_{EQ}$ bond distance does not prevail over the Mn-O$_{EQ}$-Mn angle (not
imposed by the substrate). For small thickness ($t=650$\AA ), the $T_{IM}$\
transition is favored with respect to the CO/OO\ state.

\section{Conclusion}

In conclusion, we grown high quality thin films of Nd$_{0.5}$Ca$_{0.5}$MnO$%
_3 $ on SrTiO$_3$ and LaAlO$_3$ using the PLD technique. We have shown that
we are able to induce an insulator-to-metal transition at a magnetic field
much lower than the corresponding bulk compound by the modification of the
structure through the utilization of the appropriate substrate. Moreover the
substrate-induced strains increases the Mn-O$_{EQ}$-Mn bond angles on both
LAO and STO substrates (even if the orientation of the film is different)
favoring the double exchange and consequently destabilizing the
charge/orbital state leading to a decrease of the critical magnetic field as
compared to the bulk.

\newpage

Figures Captions:

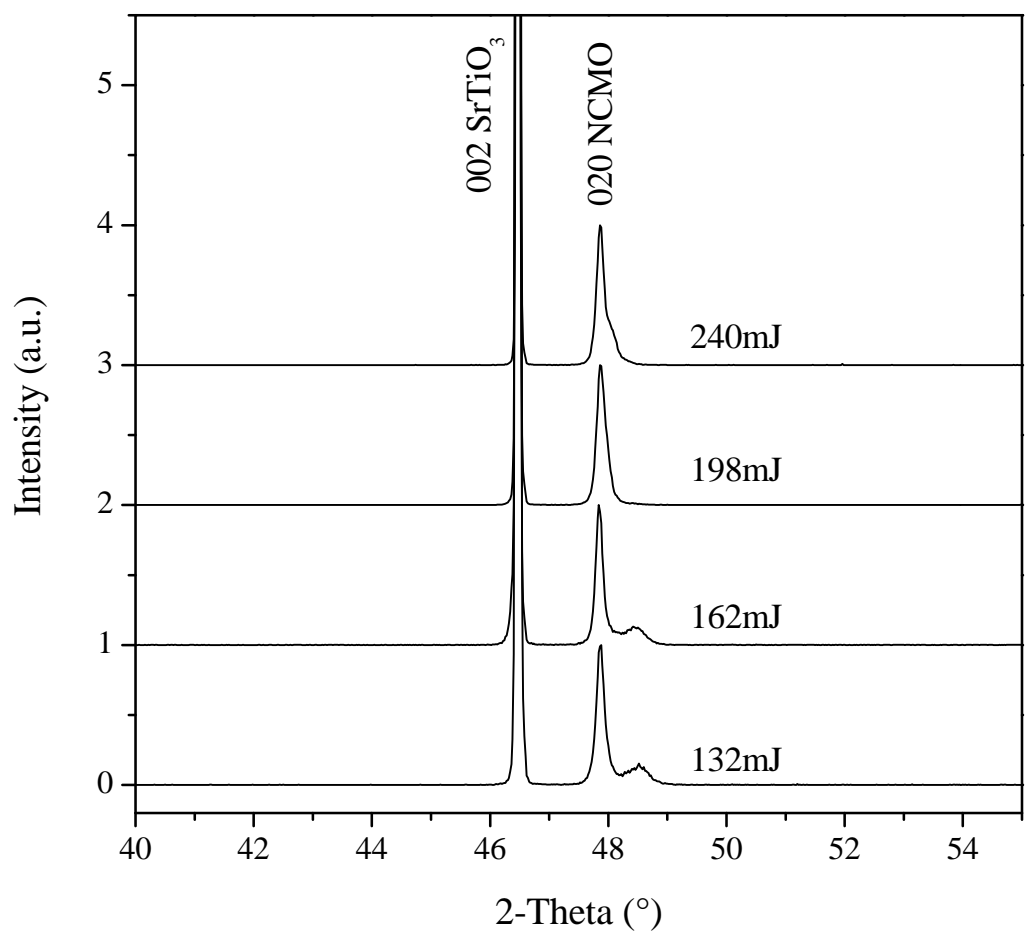
Fig.1: Room temperature XRD in the range 40-55${{}^\circ}$ of a 2000\AA\
NCMO film on STO grown with various energy of the laser beam.

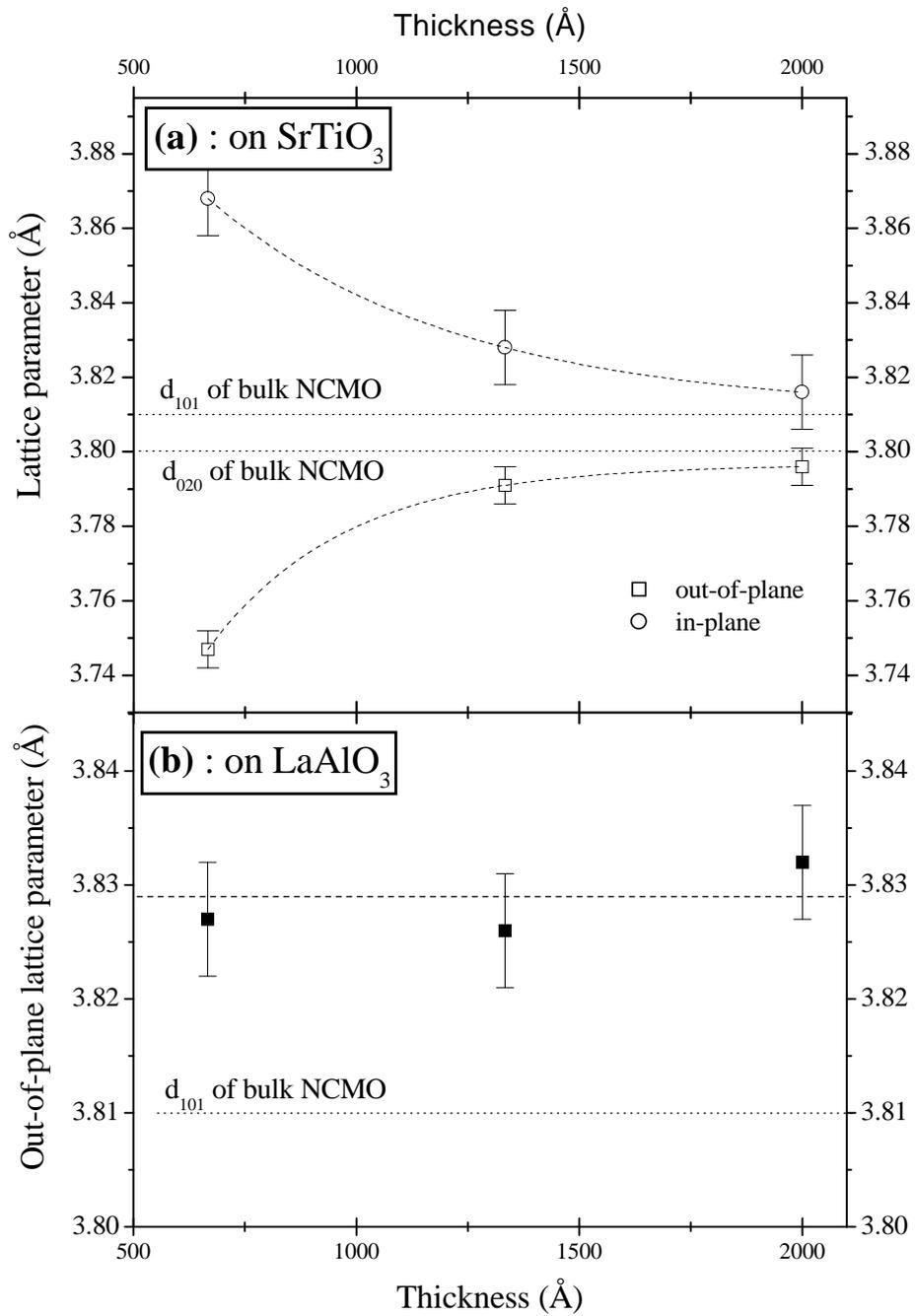
Fig.2: : (a): Evolution of the lattice parameters of NCMO on STO as a
function of the thickness. (b): Evolution of the out-of-plane lattice
parameters of NCMO on LAO as a function of the thickness The values of the
bulk are indicated as a dot line (see text for details). Lines are only
guide for the eyes.

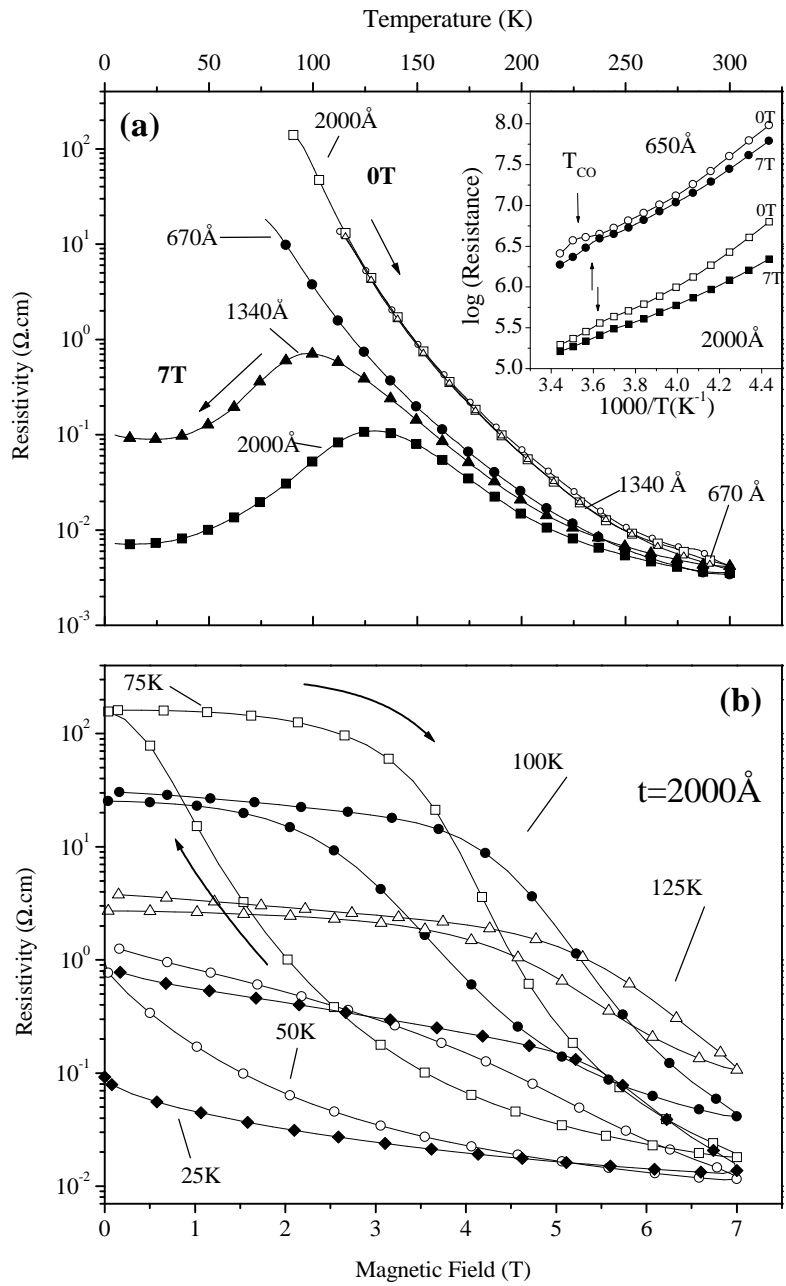
Fig.3: NCMO films on STO with various thickness. (a): $\rho (T)$ under 0T
(empty symbols) and 7T (full symbols). The inset depicts the evolution of
the logarithmic resistivity versus the inverse of the temperature under 0T
(empty symbols) and 7T (full symbols). Note the anomaly associated to the $%
T_{CO}$ (marked by arrow). (b): $\rho (H)$ at various temperature for a
2000\AA\ film.

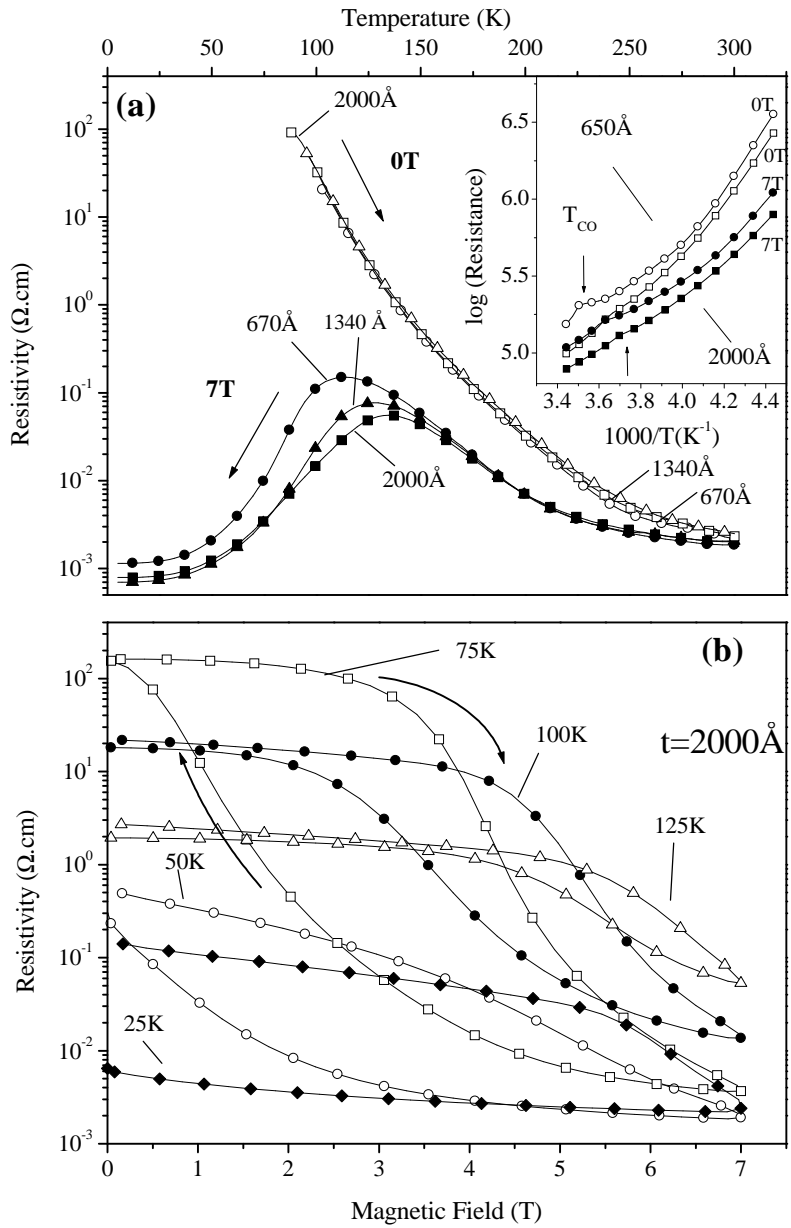
Fig.4: NCMO films on LAO with various thickness. (a): $\rho (T)$ under 0T
(empty symbols) and 7T (full symbols). The inset depicts the Evolution of
the logarithmic resistivity versus the inverse of the temperature under 0T
(empty symbols) and 7T (full symbols). Note the anomaly associated to the $%
T_{CO}$ (marked by arrow). (b): $\rho (H)$ at various temperature for a
2000\AA\ film.

Fig.5a: Idealized [010] NCMO structure on STO. The directions of the stress
are indicate by arrows. F and S refer to the film and to the substrate.

Fig.5b: Idealized [101] NCMO structure on LAO. The directions of the stress
are indicate by arrows. F and S refer to the film and to the substrate.

\end{document}